# A Study of Teacher Educators' Skill and ICT Integration in Online Teaching during the Pandemic Situation in India

Subaveerapandiyan A
*Regional Institute of Education Mysore*, subaveerapandiyan@gmail.com

R Nandhakumar
*Regional Institute of Education, Mysuru*, nandha_7025@yahoo.com

# A Study of Teacher Educators' Skill and ICT Integration in Online Teaching during the Pandemic Situation in India

Subaveerapandiyan A[1], R.Nandhakumar[2]

[1]Professional Assistant, Regional Institute of Education, Mysuru, INDIA
E-mail ID: subaveerapandiyan@gmail.com

[2]Assistant Professor (Contractual), Department of Education,
Regional Institute of Education, Mysuru, INDIA
Mobile: 9842408251, E-mail ID: nandha_7025@yahoo.com

## Abstract

Information and communication technology prompted the sharing of information over the world. For its impact on education, the government and the authorities like the University Grants Commission in India have energized the higher education institutions in India to implement online education during the pandemic situation. This paper attempts to know the teaching faculties' ICT skills and related online class skills in higher educational institutions in India. In India, like developing countries, quick as the lightning change in traditional to fully online classes are like a rumble of thunder because faculties are adopting this situation but students are challenging to adopt. The directed out of two hundred and twenty respondents were sampled randomly, questionnaires were distributed online, 201 were properly filled and returned from central universities, government higher education institutions, and state universities. A one-way ANOVA and independent-sample t-test on SPSS for data analysis were adopted. The results revealed that the respondents are significantly the same regarding their skills related to online teaching and ICT integration. Online education also involves sharing of resources, reinforcement techniques, questioning, and evaluation. There was a significant difference in ICT skills and ICT integration in online education. This was due to their age, availability of resources and some operational difficulties. Following the present scenario, the right attitude and skills are recommended on the 21st-century requirements. Faculties are how they feel and taking into serious or joy, and an online class is given to their pleasure or pressure are discussed in this article.

1. **Introduction**

Information and communication technology (ICT) is becoming the leading channel in transforming education all over the world. It helps the professional development of teachers at higher education levels. Online education is responsible for the dissemination of knowledge and information technologies. With the advent of ICT, teachers, students and institutions are prepared to meet the new challenges in the field of educational technology to produce skillful learners and accelerate learning. Online learning is the focus of this particular investigation.

Many studies have been conducted in this field, but all those studies focused on knowledge management, skills, and strategies. However, this study focused on the teacher's mentality towards online learning. Attitude plays a significant role in the implementation's success of online teaching in a mandatory situation. It involves both the teaching experience based on ICT applications and skill orientation. A professionally competent teacher will face problems in ICT integration and thus will become the pressure for the teacher. At the same time, when they are capable of overcoming any complexities in ICT integration, it becomes a pleasure for them. This study also focused on teachers' skill at higher education levels according to the needs of qualified teachers who might serve both teaching experience based on ICT applications in the classroom from one side and practical issues inside the online environment from another and managerial purposes.

The new educational policy 2021 adopted by the Indian government supports online teaching at every level, particularly at the higher education level. The country got a new education policy after 34 years. The new education policy is framed with the vision of creating a higher education system that contributes directly to transforming the country, providing high-quality education to all, and making India a global knowledge superpower. The new educational policy emphasizes the need to use technology in teaching, learning and assessment, education planning, digital India campaign, administration and management and regulations through public disclosures. National Educational Technology Forum (NETF), MOOCs, Divyang friendly educational software, e-content in various languages, to meet 21st-century challenges, virtual labs, and online assessment and examination are some of the products of technology in education based on the new policy of education. In this way, in the field of education, the government has highly significantly prioritized ICT development with the mission that teachers at the higher education levels should be ready for such interaction with the government to work together to overcome such issues. Based on this point of view, the existence of ICT integration in higher education refers to the understanding of ICT and its impact. While it might be integrated with a great interest in the curriculum by the students, teacher capacity building as well as the availability of resources like computers with network connectivity, at this time is lacking. This is the primary reason for pressure in the minds of the teachers. Many higher education institutions are provided with 24 hours free wi-fi connectivity, adequate computer systems, laptops and tablets for the teachers and students in recent years. Within these circumstances, all the general curriculum textbooks are converted into digital textbooks available online. Computers should cover all the higher education systems in India as per the new policy of education.

Based on the survey among the teachers in higher education institutions, in the academic year 2020-2021, it was the first time the student at higher education level taught utterly in an online mode as per the policy of the University Grants Commission(UGC), India but has not been monitored thoroughly. Perhaps these refer to the case of many barriers, such as lack of computer devices and skilled teachers. From this point of view, when online courses are implemented in a compulsory manner, the teachers have a chance to get some experience of using online resources with regards to their student's learning. In spite of all difficulties, teachers felt pressure, but at the same time, the uninterrupted service made pleasure in the minds of the teachers. Consequently, more training needs to be conducted for teachers in order to develop skills related to online education.

### 1.1 The impact of ICT use on teaching

ICT within educations is increasing with the demands because of the political, social and economic conditions of a society. Therefore, ICT becomes an integrated system globally to speed up the knowledge and abilities of students. Teachers progressively increase information when the classroom environments are ready for ICT integration. (Law, 2000). The classroom infrastructure, technical advancements, equipment and human resources such as teachers, technical administrators and end-users are the components of this system. Presentation in online teaching could be managed to get up-to-date knowledge and information that will refer to both teachers and students and could improve school management system instead of traditional teaching methods.

**1.2 Teacher's ICT skills towards online teaching**

Shamim and Raihan, 2000 conducted a study of the effectiveness of ICT in the online teaching environment. The results revealed that ICT in online teaching is valuable and purposeful due to time-saving and versatility. In order to infuse ICT in the online teaching environment, academic knowledge and managing skills, combining a wide range of skill sets are necessary. Utilizing technology in the online classroom could assist the teacher in facilitating teaching, which can lead to understanding the skills that teachers using online teaching possessed, which help enhance student's learning are. ICT needs to be incorporated into the curriculum by effective teachers who make learning to interest enjoyable, motivate students, and practice skills they will need in their everyday work lives. (Tella, Toyoba, Adika and Adewuyi, 2007). When there is pleasure in teachers' minds regarding online teaching and usage within the classroom, hopefully, interested teachers will reform the system to provide students with the knowledge and skill to ensure successful implementation. On the other hand, when the skills were low among the teachers, they feel it was a working pressure. Alazam, Bakar, Hamzah and Asmiran, 2012 conducted a study regarding teacher's skill and ICT integration with Online teaching in Malaysia, and the results revealed that there is a significant relationship between ICT use and teacher's skill. The skills were low among teachers and the disparities demographically due to age, and this confirmed that there is a relationship between age and teacher's skill. Teacher's beliefs were signified as one of the factors that ICT integration all over the world (Tsai & Chai, 2012) and teaching experience (Boyd, Lankford, Loeb, Rokoff & Wyckoff, 2008). Teachers require extensive exposure to online teaching. This paper examined teacher's skills at the higher education level in Indian higher education institutions.

**2. Methodology**

The online questionnaire is the research tool used for this study for data collection from online education. The teachers were employees of Central universities, State universities and National institutes run by the ministries of the Government of India.

**2.1 Sampling**

The population of this study consists of teachers who are teaching content and pedagogical subjects in education. Teachers were selected as samples randomly from 26 central universities, 43 state universities, 6 national institutions, and 2 other institutions in India, consisting of Departments of Education, Educational technology, Special education, and Continuing education. Teachers are in the same characteristics.

**Demographic Information:**

**Table – 1 Demographics characteristics of Respondents (N= 201)**

| Variables | Categories | n (%) |
|---|---|---|
| Gender | Female | 118 (58.7%) |
| | Male | 83 (41.3%) |
| Age (Years) | Below 27 | 4 (2%) |
| | 27-35 | 26 (13%) |
| | 35-49 | 123 (61.5%) |
| | 50 and Above | 47 (23.5%) |
| Academic Rank | Assistant Professor | 131 (65.2%) |
| | Associate Professor | 19 (9.4%) |
| | Professor | 41 (20.4%) |
| | Other | 10 (5%) |
| Experience | 1-5 Years | 30 (14.9%) |
| | 6-10 Years | 53 (26.4%) |
| | 11-20 Years | 78 (38.8%) |
| | Over 20 Years | 40 (19.9%) |

## 2.2 Instrumentation

The questionnaire was developed by the author -1 in consultation with the teachers working in the Regional Institute of Education, Mysuru. The questionnaire was made available in online mode. The questionnaire has two parts; part-1 contained the teacher's demographic information such as type of the organization, gender, age, academic rank, years of teaching experience, and part – 2 holds 8 structured items on variables to gather the qualitative data from the teachers. These variables are classified into the mode of delivery during the online teaching, satisfaction level, technical issues, challenges in designing the content, enriching the content, factors enriching online teaching, enjoying teaching and opinion online teaching. Besides these, information about their computer skill, devices used for online teaching and the platform preferred was also collected. 5- point Likert scale (1–Strongly Disagree 2-Disagree 3-Not sure 4–Agree 5–Strongly agree) has been used to ascertain the skill and exposure of respondents.

**Figure 1. Organization Type**

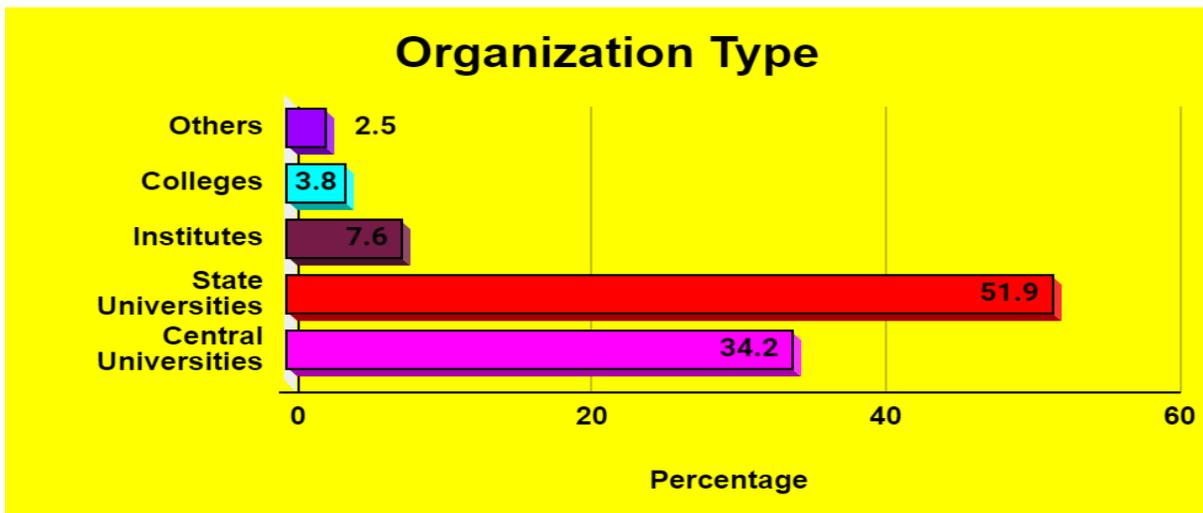

The Figure 1 shows the percentage of respondents for the investigations. Teachers working in State universities responded well with 51.9%, followed by Central universities with 34.2% and Institutes with a 7.6% Percentage. Teachers Working colleges respondents 3.8%.

**Figure 2. Gender**

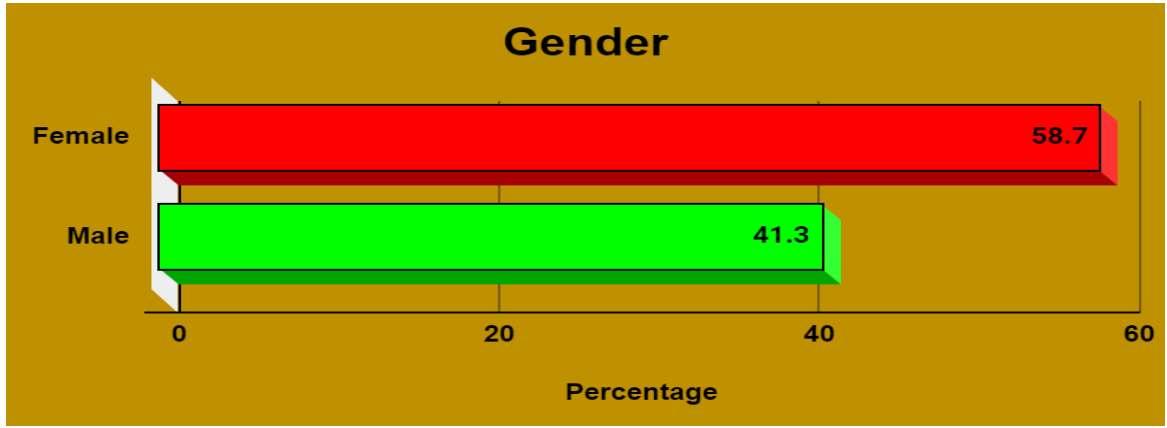

Figure 2 represents the gender-wise distribution where 58.7% were female teachers and 41.3% were male.

**Figure 3. Age**

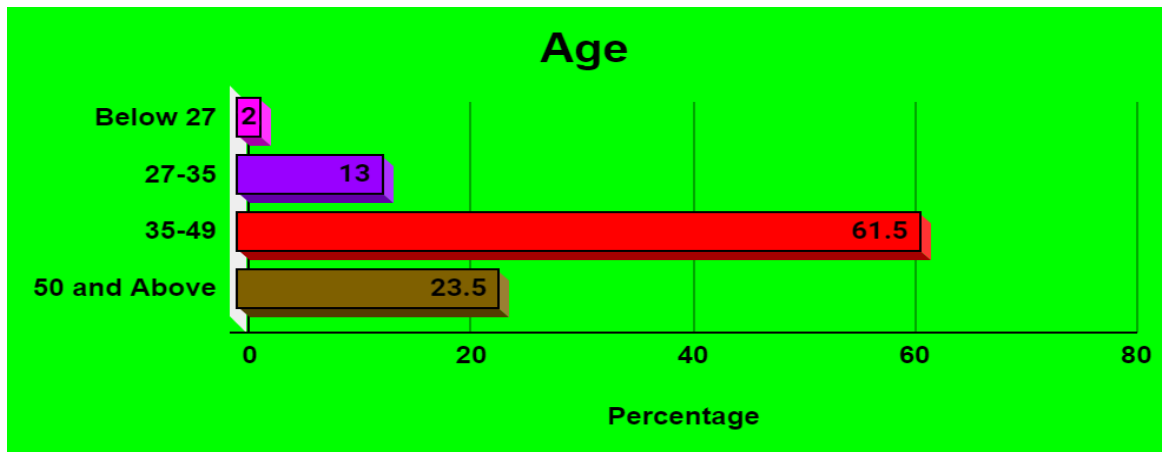

Figure 3 provide details of the age-wise respondents. The highest respondents are 35 to 49 years are 61.5%, the lowest respondents are below 27 is 2% of respondents.

**Figure 4. Academic Rank**

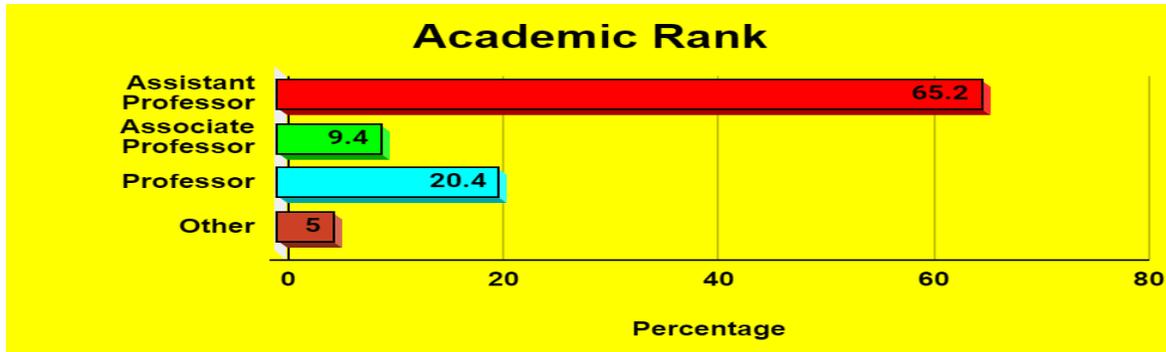

Figure 4 represents the academic rank of the faculty members. The Associate professors are 65.2% is the highest in percentage, followed by the Professors are 20.4%, Associate professors are 9.5% and the last and fewest respondents are Others is 5%.

**Figure 5. Teaching Experience**

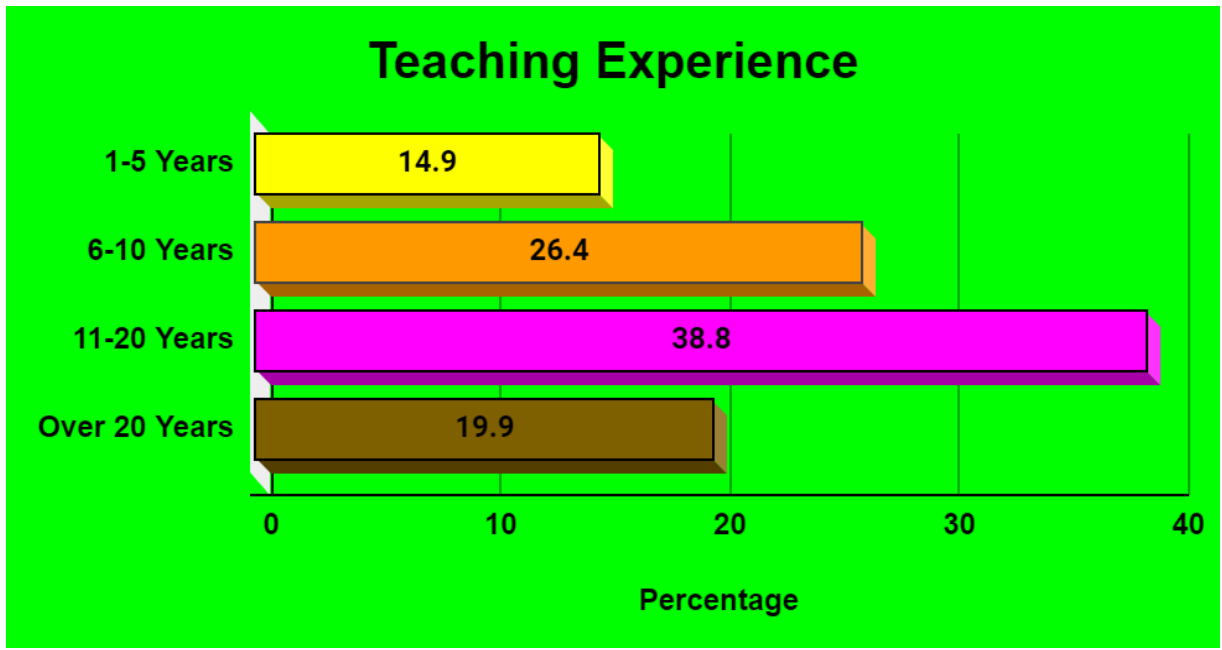

Figure 5 represents the teaching experience of the faculty members, the 11 to 20 year's experience is 38.8% is the highest in percentage, followed by the 6 to 10 years are 26.4%, Over 20 years' experience are 19.9% and last and 1 to 5 years' experience respondents 14.9%.

**Table -2 Computer Competency**

| Computer Competency | Percentage |
|---|---|
| Beginner | 4.5 |
| Intermediate | 76.6 |
| Expert | 18.9 |

Table 2 shows the computer competency of the faculty members. Here it clearly presents 76.6% in the intermediate level, 18.9% respondents are only expert and 4.5% are beginner level.

**Table -3 Online Platform Preferred for Teaching**

| Online platform preferred | Percentage |
|---|---|
| Zoom | 30.8 |
| Google Classroom | 62.7 |
| WhatsApp | 0.5 |
| Gotowebinar | 2.5 |
| Other | 3.5 |

Table 3 shows an online platform preferred for teaching. It used the highest number of respondents Google Classroom is 62.7%, the lowest usage platform is WhatsApp 0.5% of respondents.

**Table -4 Device (s) Used for Online Teaching**

| Device (s) used for your online teaching | Percentage |
|---|---|
| Desktop | 8 |
| Laptop | 67.1 |
| Tablet | 1 |
| Smart Phone | 22.4 |
| All of the above | 1.5 |

The above table reveals the details of the device used for online teaching respondents. The highest respondents are used laptop 67.1%, the lowest respondents are used tablet 1% of respondents.

## 3. Results

### 3.1 Research question–1: Differences between teachers' ICT skill in online teaching due to their attitude in the teachers' mode of delivery and satisfaction level during online teaching and learning

The study examines the differences in ICT skills in online teaching due to their attitude in the mode of delivery during online teaching and learning and the satisfaction level of the teachers during online teaching. To analyze this the percentage of responses was analyzed as per the Table-5. The majority of respondents felt interesting and few teachers are felt boring. Similarly, the satisfaction level of the teachers with online teaching is high as per the Table – 6.

**Table -5 The attitude in the mode of delivery during online teaching and learning.**

| The attitude in the mode of delivery during online teaching and learning is | Percentage |
|---|---|
| Interesting | 47.8 |
| Intimidating | 38.8 |
| Boring | 13.4 |

**Table -6 Satisfied with the technology and software you are using for online teaching**

| Satisfaction Level | Percentage |
|---|---|
| High | 62.2 |
| Low | 10.4 |
| Moderate | 27.4 |

### 3.2 Research question–2: the technical issues related to the transition of online classes

To answer this question, for the analysis procedure, the percentage analysis was used. The highest technical issue among the teachers were presented in the Table – 7.

**Table -7 Technical issues related to the transition of online classes**

| Technical issues related to the transition of online classes | Percentage |
|---|---|
| Student discomfort or lack of familiarity with required technologies or applications | 67.7 |
| My own discomfort or lack of familiarity with required technologies or applications | 7 |
| My access to reliable communication software/tools (e.g. Zoom, Skype, Google) | 11 |
| My access to reliable internet/service | 15.4 |
| My access to a reliable digital device (e.g. laptop, mobile device) | 5.5 |
| My access to specialized software (e.g. Adobe products, statistical packages) | 10 |
| My access to library resources | 8.4 |
| Other | 4 |

**3.3 Research Question 3: The level of challenges of teachers in designing online content**

To answer this question, for the analysis procedure the percentage of response is studied. As shown in the Table 8, the level of challenges of teachers in designing online content is studied. The highest challenge has been "Personal preference is Face-to-Face learning" followed by "Lack of personal contact with the students".

**Table -8 Challenging on designing online content**

| Challenging on designing online content | Percentage |
|---|---|
| I am not familiar or comfortable with online applications/tools | 24.9 |
| I have limited knowledge of options for online course delivery | 19.9 |
| My personal preference is for face-to-face learning | 53.7 |
| Getting adequate support to manage my transition to online learning | 25.3 |
| Lack of personal contact with my students | 50.7 |
| The actual teaching online | 15.4 |
| Course lessons or activities have not translated well to a remote environment | 27.3 |
| I am uncertain about how to best assess student learning in this environment | 34.3 |
| Students have not been adequately available/responsive | 36.3 |
| Others | 2.5 |

**3.4 Research Question 4: the factors enriching online teaching.**

As shown in the Table 9, the factors enriching online teaching is studied. "flexibility is the highest preference" followed by "innovation" and "Usefulness" was the third factor ranked.

**Table -9 Factors enrich online teaching**

| Factors enrich online teaching | Percentage |
|---|---|
| Flexibility | 64.7 |
| Wide Range of Tools | 36.8 |
| Ease of Use | 38.8 |
| Usefulness | 42.8 |
| Customization (ability to personalize learning for students) | 30.3 |
| Innovation (i.e. freedom to experiment with teaching practice) | 50.2 |
| Accessibility(Platforms, Materials, Resources) | 39.8 |
| Engagement and enjoyment of pupils | 27.9 |
| Nothing | 10 |
| I haven't had any experience with online/distance learning | 5.4 |
| Other | 0 |

**3.5 Research question -5 Are you enjoying teaching your students remotely and do you have Pleasure or Pressure**

It is clear from the table - 10 and figure–6 that 50.7% of teachers are enjoying teaching through online mode and 22.4% responded No. When asked about 'Pleasure' or 'Pressure', most of the respondents (71.6%) answered both pleasure and pressure as per the table – 11.

**Table -10 Enjoying teaching your students remotely**

| Enjoying teaching your students remotely | Percentage |
|---|---|
| Yes | 50.7 |
| No | 22.4 |
| Maybe | 26.9 |

**Figure 6. Are you enjoying teaching your students remotely?**

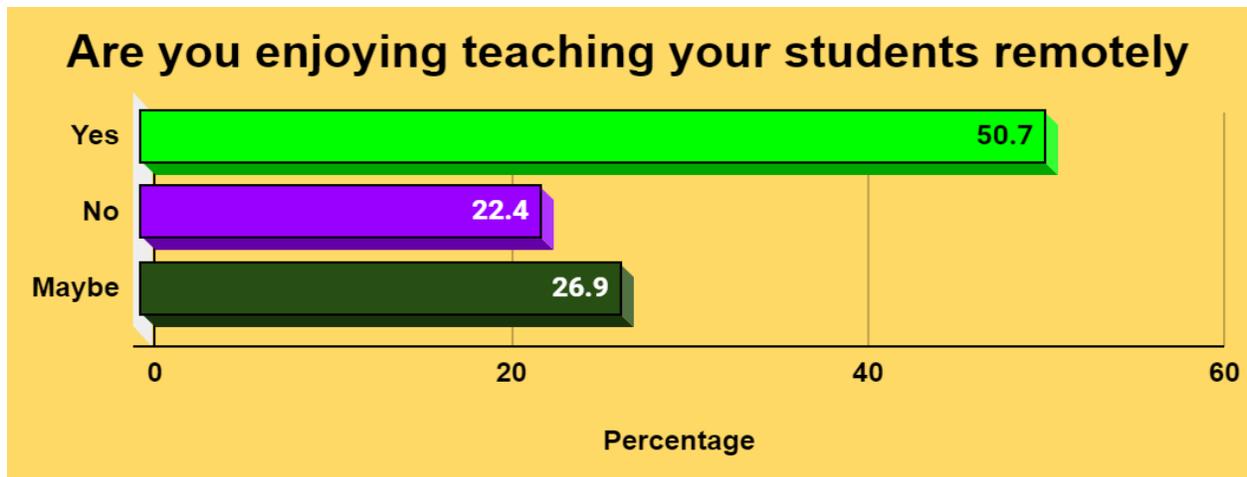

**Table -11 Opinion on online teaching**

| Opinion on online teaching | Percentage |
|---|---|
| Pleasure | 20.9 |
| Pressure | 7.5 |
| Both | 71.6 |

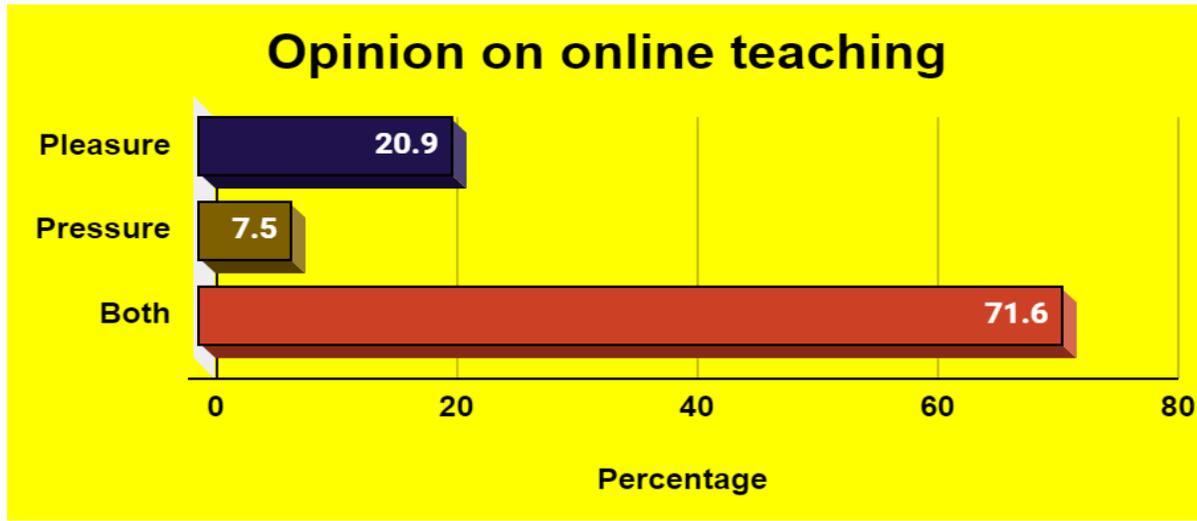

**Figure 7. Opinion on online teaching**

### 4. Analysis: Online teaching – A pleasure or pressure

To examine whether the online teaching is pleasure or pressure, the differences in attitude on male and female was studied and an independent T-test was conducted as presented in Table 11. The table given below shows that there was a statistically significant difference reported in the attitude among teachers. Males and females were different scores in pleasure. (M = 28.17; S.D. = 7.17) in male and female (M = 24.73; S.D. = 6.87); t (202) = 3.14, P = 0.00036. Males and females were different scores in the pressure. Male (M = 11.86; S.D. = 7.2) and female (M = 15.3; S.D. = 6.9); t (202) = -3.42, P = 0.000381.

**Table – 12 Pleasure or Pressure**

| Pleasure | N | DF | Mean | SD | T Value | P-value |
|---|---|---|---|---|---|---|
| Male | 83 | 82 | 28.17 | 7.17 | 3.44 | .00036. significant |
| Female | 119 | 118 | 24.73 | 6.87 | | |

| Pressure | N | DF | Mean | SD | T Value | P-value |
|---|---|---|---|---|---|---|
| Male | 83 | 82 | 11.86 | 7.2 | -3.42 | .000381 significant |
| Female | 119 | 118 | 15.3 | 6.94 | | |

Table 12 shows that there was statistically significant difference reported in both pressure and pleasure. Although, the experience ranged between 1-5 year scored higher mean (M = 28.4; S.D. = 5.17 ) in pleasure, experience over 20 years scored the lowest mean. (M=25.05; SD= 8.03). The experience ranged between 6-10 years scored higher mean in Pressure (M= 15.46; SD=9.11), and experience between 1-5 years scores lower mean.(M-11.6; M= 5.17).

**Table-12 Pleasure or Pressure with respect to experience**

| Experience | | N | DF | Mean | SD | T Value | P-value |
|---|---|---|---|---|---|---|---|
| 1-5 years | Pleasure | 30 | 29 | 28.4 | 5.17 | 12.58482 | .00001. significant |
| | Pressure | | | 11.6 | 5.17 | | |
| 6-10 Years | Pleasure | 54 | 53 | 24.61 | 9 | 5.24629 | .00001. significant |
| | Pressure | | | 15.46 | 9.11 | | |
| 11-20 Years | Pleasure | 77 | 76 | 27.13 | 5.42 | 16.19963 | .00001. significant |
| | Pressure | | | 12.9 | 5.47 | | |
| Over 20 years | Pleasure | 41 | 40 | 25.05 | 8.03 | 5.69054 | .00001. significant |
| | Pressure | | | 14.95 | 8.03 | | |

The pleasure or pressure due to the occupational stress is analyzed. Table 12 shows that there is no significant difference between male and female in their attitude towards online teaching due to the occupational stress and there is no significant difference in attitude at all the levels of teaching experience as per table – 13.

**Table 13 Pleasure or Pressure due to the occupational stress**

| Pleasure | N | DF | Mean | SD | T Value | P-value |
|---|---|---|---|---|---|---|
| Male | 83 | 82 | 15.99 | 6.11 | 1.01 | .156979 not significant |
| Female | 119 | 118 | 15.05 | 6.74 | | |

| Pressure | N | DF | Mean | SD | T Value | P-value |
|---|---|---|---|---|---|---|
| Male | 83 | 82 | 16.01 | 6.11 | -1.01 | .156979 not significant |
| Female | 119 | 118 | 16.95 | 6.74 | | |

## 5. Conclusion

From the study results, it was found that based on the teachers' demographic information, the teachers in the age group of 35-49 were more skilful when compared with their colleagues. In addition, computer competency is intermediate among the majority of teachers use it. The Google platform and Zoom online platform are preferred as major online platform. And only the laptop had been used among the respondents rather than the mobile phone for teaching online, regarding the subjects and materials which might be used for their students. Based on our findings, the authors strongly recommended that, as a priority work should be taken into account is that; long-term of the strategic plan for preparing online teaching with enough computer devices and internet connectivity, ICT training should be integrated to enable pre-service teachers to acquire the necessary skills needed for the online teaching.

Information literacy, media literacy, and information and communication technologies are being strongly recommended to be included in the teacher-training program. Thus, the teachers can be able to use the ICT technology anywhere and anytime for their students' benefit in the presence of all the aids of practice. For further research in this study, we recommend studying the teacher professional development (TPD) based on ICT skills, evaluate teachers' technological pedagogical and content knowledge (TPACK).